\documentclass[letter]{aa} 

\usepackage{graphicx}
\usepackage{txfonts}
\usepackage[colorlinks=true, allcolors=blue]{hyperref}

\usepackage{color}

\begin{document}

\title{Spectroscopic confirmation of the galaxy clusters CARLA J0950+2743 at z=2.363, and CARLA-Ser J0950+2743 at z=2.243}
\titlerunning{Spectroscopic confirmation of CARLA J0950+2743 and CARLA-Ser J0950+2743}
\authorrunning{Grishin et al.}

\author{Kirill A. Grishin\inst{1}
          \and
          Simona Mei\inst{1,2}
           \and
          Igor V. Chilingarian\inst{3,4}
          \and
          Marika Lepore\inst{5}
          \and
          Paolo Tozzi\inst{5}
          \and
          Anthony Gonzalez\inst{6}
          \and
          Nina Hatch\inst{7}
          \and
          Spencer A. Stanford\inst{8}
          \and 
          Dominika Wylezalek\inst{9}
          }

\institute{Universit\'e Paris Cit\'e, CNRS(/IN2P3), Astroparticule et Cosmologie, F-75013 Paris, France \email{grishin@voxastro.org}
\and Jet Propulsion Laboratory and Cahill Center for Astronomy \& Astrophysics, California Institute of Technology, 4800 Oak Grove Drive, Pasadena, California 91011, USA
\and
Center for Astrophysics --- Harvard and Smithsonian, 60 Garden Street MS09, Cambridge, MA 02138, USA
\and 
Sternberg Astronomical Institute, Moscow M.V. Lomonosov State University, Universitetsky pr., 13,  Moscow, 119234, Russia
\and
INAF-Osservatorio Astrofisico di Arcetri, Largo Enrico Fermi 5, 50125, Florence, Italy
\and
Department of Astronomy, University of Florida, 211 Bryant Space Center, Gainesville, FL 32611, USA
\and
School of Physics and Astronomy, University of Nottingham, University Park, Nottingham NG7 2RD, UK
\and
 Department of Physics, University of California, One Shields Avenue, Davis, CA 95616, USA
 \and
Astronomisches Rechen-Institut, Zentrum fur Astronomie der Universitat Heidelberg, Monchhofstr. 12-14, 69120, Heidelberg, Germany
}

\date{Received ; accepted }

\abstract{
Galaxy clusters, being the largest gravitationally bound structures in the Universe, are a powerful tool to study mass assembly at different epochs. At z$>$2 they give an unique opportunity to put solid constraints not only on dark matter halo growth, but also on the mechanisms of galaxy quenching and morphological transformation when the Universe was younger than 3.3 Gyr. However, the currently available sample of confirmed $z>2$ clusters remains very limited. We present the spectroscopic confirmation of the galaxy cluster CARLA~J0950+2743 at $z=2.363\pm0.005$ and a new serendipitously discovered cluster, CARLA-Ser J0950+2743 at $z=2.243\pm0.008$ in the same region. We confirm eight star-forming galaxies in the first cluster, and five in the second by detecting [O{\sc ii}], [O{\sc iii}] and $H\alpha$ emission lines.  The analysis of a serendipitous X-ray observation of this field from Chandra reveals a counterpart with a total luminosity of $L_{0.5-5 keV} = 2.9\pm0.6\times10^{45}$~erg~s$^{-1}$. Given the limited depth of the X-ray observations, we cannot distinguish the 1-D profile of the source from a PSF model, however, our statistical analysis of the 2-D profile favors an extended component that could be associated to a thermal contribution from the intra-cluster medium (ICM). If the extended X-ray emission is due to the hot ICM, the total dark matter mass for the two clusters would be $M_{200}=3.30 ^{+0.23}_{-0.26 (\mathrm{stat})}$ $^{+1.28}_{-0.96 (\mathrm{sys})} \times10^{14} M_{\odot}$. This makes our two clusters interesting targets for studies of the structure growth in the cosmological context. However, future investigations would require deeper high-resolution X-ray and spectroscopic observations.
}


\maketitle
%

\section{Introduction}

Galaxy clusters are the largest gravitationally bound structures in the Universe, which makes them direct probes of not only the structure growth, but also of the environmentally driven galaxy evolution and transformation processes~\citep{2021NatAs...5.1308G}. Galaxy cluster studies have important cosmological implications provided that they can constrain baryon processes and mass assembly channels at different epochs~\citep[e.g., ][]{2011A&A...531L..15S, 2014ApJ...796...65M, 2019A&A...622A.117S, 2023A&A...670A..58M, 2023A&A...670A..95A, 2023arXiv231016085K, 2024arXiv240208458G}.

The upcoming deep wide-field sky surveys such as the Rubin Legacy Survey of Space and Time~\citep[LSST; ][]{2018cosp...42E1651K}, Euclid~\citep{Euclid-r}, and the Nancy Grace Roman Telescope~\citep{2021MNRAS.507.1746E} as well as  sub-mm surveys like CMB-S4~\citep{2016arXiv161002743A}, will provide the opportunity to systematically study properties of galaxy clusters and their individual members out to $z\sim2-3$~\citep[e.g., ][]{2015MNRAS.453.2515A}.

However, the currently available sample of known galaxy clusters at $z>2$ is too small to understand their statistical properties crucial for the preparation for future surveys. 
Deep observations using Hubble and James Webb Space Telescopes opened new discovery space for galaxy clusters and proto-clusters at z$>$2~\citep[e.g., ][]{2016ApJ...828...56W, 2018ApJ...859...38N, 2022A&A...667L...3L, 2023ApJ...947L..24M}. At the same time, ground-based facilities still play an important role in the identification and confirmation of candidate clusters and proto-clusters~\citep{2018ApJ...862...64S, 2021MNRAS.508.3754W, 2024ApJ...970...59T, 2024arXiv240415910T,2022ApJ...930..102Y, 2024ApJ...963...49K, 2024A&A...684A.196Z}. A systematic search in the area of the COSMOS UltraVISTA field, which has a deep multi-wavelength coverage, also yielded several high-fidelity proto-clusters at $2<z<4$~\citep{2011Natur.470..233C, 2012ApJ...748L..21S, 2014ApJ...795L..20Y, 2021MNRAS.503L...1K, 2020ApJ...892....8D, 2016ApJ...817..160L, 2015ApJ...802...31D, 2015ApJ...808L..33C, 2016ApJ...828...56W, 2015ApJ...808L..33C, 2018A&A...615A..77L, 2018ApJ...861...43P, 2022ApJ...926...37M}. 

Of those detections, only two have been confirmed as clusters with intra-cluster medium (ICM) at $z>2$: $J1449-0856$ at $z=2.07$~\citep{2011A&A...526A.133G} and $J1001+0220$ at $z=2.51$~\citep{2016ApJ...828...56W}. 
 
In this letter, we present (i) the spectroscopic confirmation of a cluster candidate identified by the Clusters Around Radio-Loud Active Galactic Nuclei (AGN) survey \citep[CARLA][]{2013ApJ...769...79W,2014ApJ...786...17W}, CARLA~J0950+2743 at $z=2.363\pm0.005$, and (ii) an identification of a new cluster, CARLA-Ser~J0950+2743 at $z=2.243\pm0.008$ in the same area of the sky, and superposed to the first. 
An archival Chandra dataset reveals an extended X-ray counterpart that is consistent with the existence of ICM either in the form of hot X-ray emitting gas or as a result of inverse Compton scattering of the cosmic microwave background on radio-lobes of a radio-loud quasar. 
Throughout this paper, we adopt the $\Lambda CDM$ cosmology, with $\Omega_M  =0.3$, $\Omega_{\Lambda} =0.7$, $h=0.72$, and $\sigma_8 = 0.8$. In our X-ray analysis, we use the widely-used self-similar evolution with $E^2(z)$, which is also consistent with observations~\citep{2009ApJ...692.1033V}.  All magnitudes are given in the AB system~\citep{1983ApJ...266..713O}. 


\begin{figure}
\centering
\includegraphics[width=\hsize]{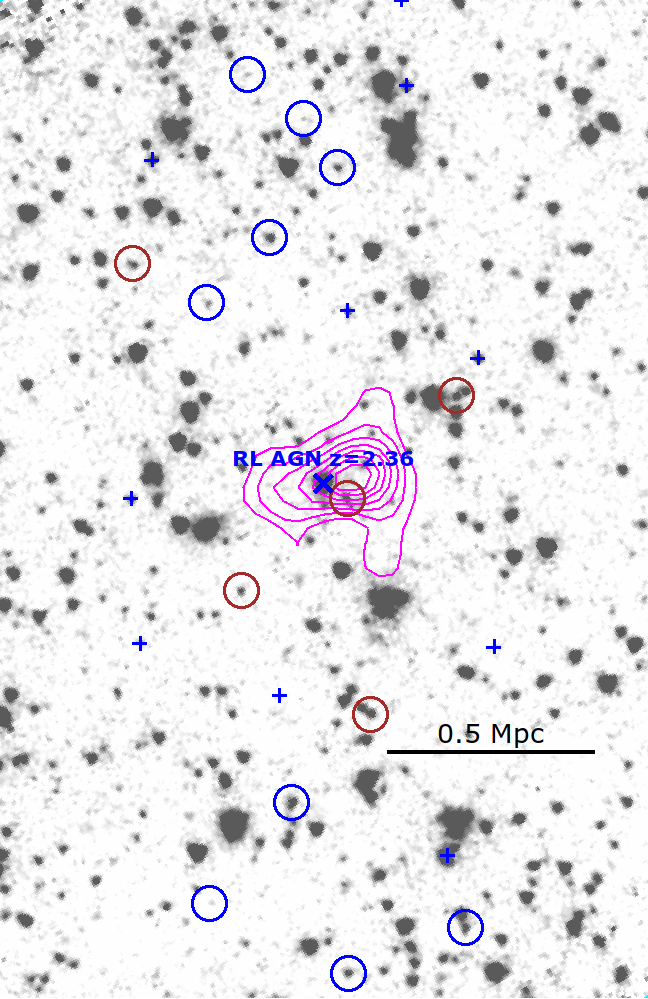}
\caption{
Spectroscopically confirmed members in CARLA J09504+2743 and CARLA-Ser J09504+2743 and X-ray counterpart (magenta contours). The blue and brown red circles denote the positions of the spectroscopically confirmed members in CARLA J09504+2743 and CARLA-Ser J09504+2743, respectively, on the \emph{Spitzer} IRAC1 image. Blue crosses show observed galaxies without detected emission lines, hence, measured redshift.
}
\label{fig_j0950}
\end{figure}

\begin{figure}
\centering
\includegraphics[width=\hsize]{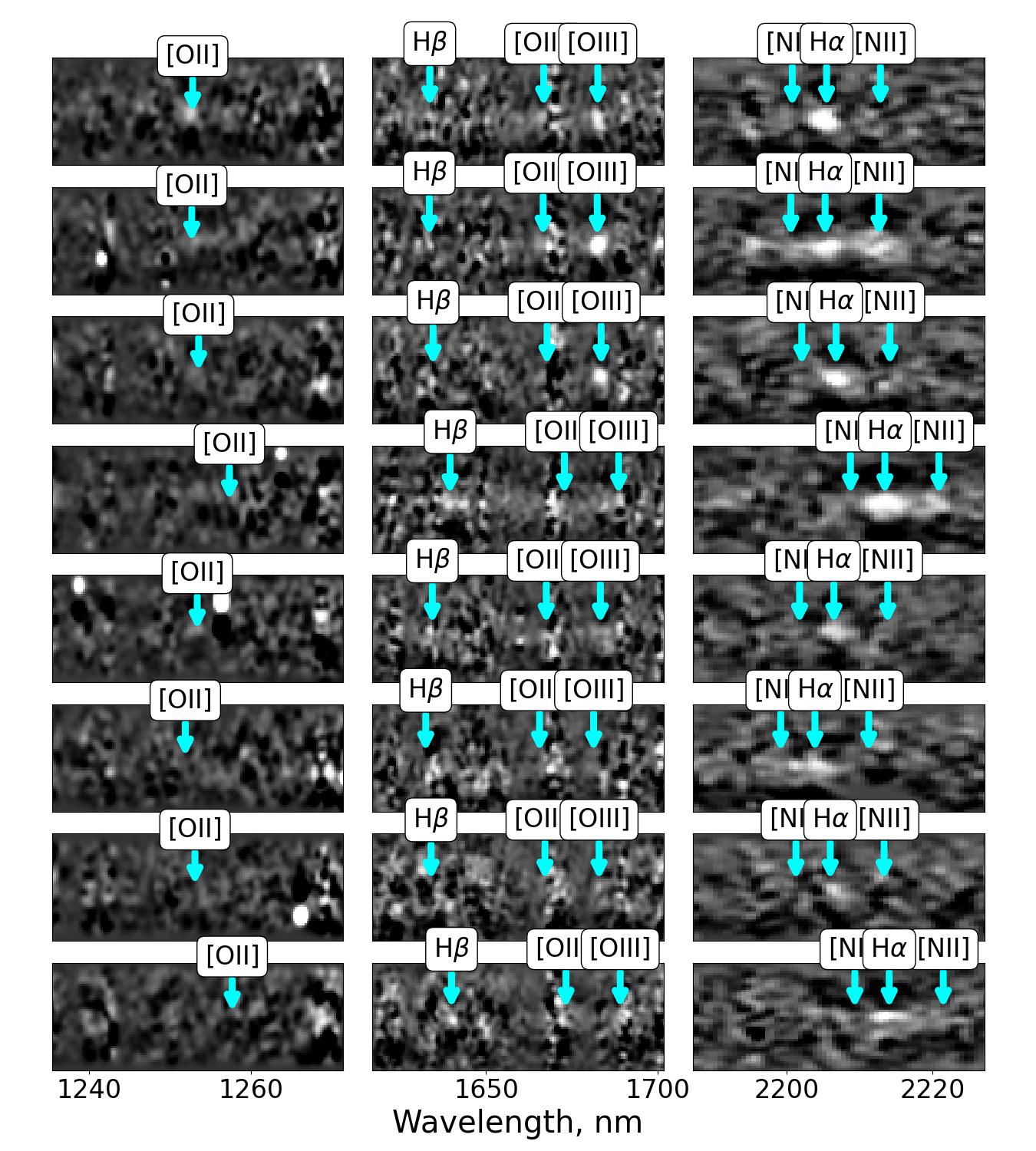}
\includegraphics[width=\hsize]{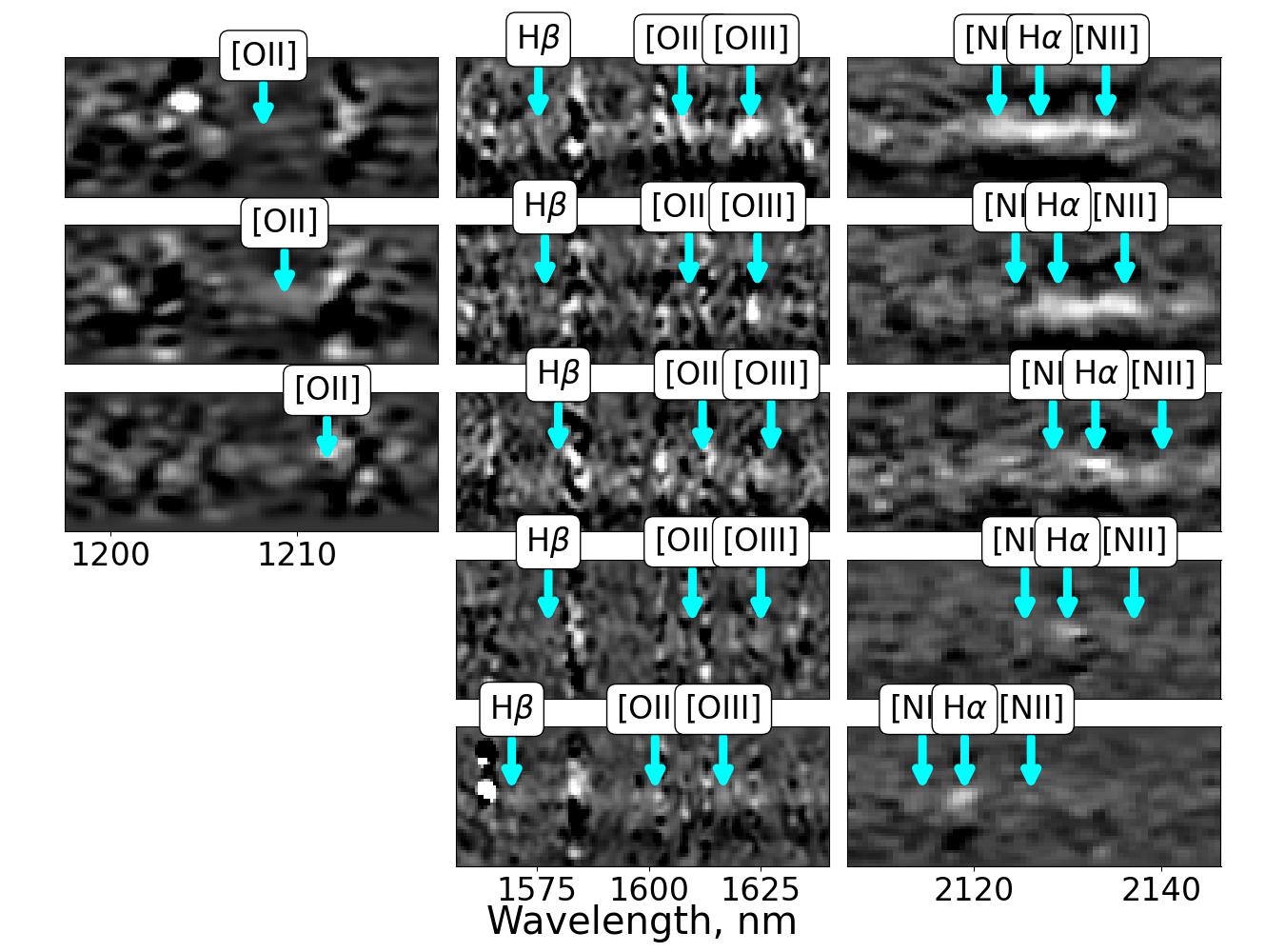}
\caption{
Parts of reduced co-added 2-D MMIRS spectra for the eight CARLA~J0950+2743 at $z=2.363\pm0.005$ (top) and the five CARLA-Ser~J0950+2743 at $z=2.243\pm0.008$ (bottom) spectroscopically confirmed cluster members. Cyan arrows indicate the positions of redshifted H$\alpha$, H$\beta$, [O{\sc iii}], and [O{\sc ii}] emission lines.  }
\label{fig_j0950_spec}
\end{figure}

\section{Observations}

\subsection{Existing Spitzer Space Telescope and ground-based observations}

CARLA J0950+2743 is a cluster candidate around an AGN at $z=2.36$ from the CARLA survey \citep[PI: D. Stern, Prop. ID: 80154; ][]{2013ApJ...769...79W,2014ApJ...786...17W}, whose main goal was the identification of galaxy cluster candidates around radio-loud AGN at $z>1.3$ by selecting galaxy overdensities with {\it Spitzer} IRAC 3.6$\mu$m (hereafter IRAC1) - IRAC 4.5$\mu$m (hereafter IRAC2) color using following criteria: 
(IRAC1-IRAC2)$>-0.1$ \citep{2013ApJ...769...79W,2014ApJ...786...17W}. 46\% and 11\% of the CARLA fields are overdense at a 2$\sigma$ and a 5$\sigma$ levels respectively, with respect to the field surface density of NIR sources in the UKIDSS Ultra Deep Survey \citep{2013ApJ...769...79W}. 

CARLA J0950+2743 shows a galaxy overdensity at $\gtrsim 3\sigma$ with respect to the field \citep{2013ApJ...769...79W,2014ApJ...786...17W}. The {\it Spitzer} IRAC1 and IRAC2 images were obtained over a common $5.2 \times 5.2$~arcmin$^2$ field of view with total exposure times of 1000~s and 2100~s, point spread function (PSF) FWHMs of 1.95~\arcsec and 2.02~\arcsec, yielding 95\%\ completeness limits of 22.6~mag and 22.9~mag, respectively \citep{2014ApJ...786...17W}.

This cluster candidate was also observed in the $i'$-band with the ACAM camera at the 4.2~m William Herschel Telescope with a total integration of 7200~sec (PI: N. Hatch). The image has a sampling of 0.25~\arcsec pix$^{-1}$, and the atmospheric seeing of 1.31~\arcsec FWHM. The 5-$\sigma$ $i$-band limiting magnitude is 24.92~mag \citep{2015MNRAS.452.2318C}.

The CARLA J0950+2743 AGN was spectroscopically observed three times by the Sloan Digital Sky Survey~\citep[SDSS; ][]{2009ApJS..182..543A} and Extended Baryon Oscillation Spectroscopic Survey~\citep[eBOSS; ][]{2020ApJS..249....3A}. The average redshift measurement is $z_{AGN} = 2.354 \pm 0.004$ based on measurements from one SDSS~\citep{2009ApJS..182..543A} and two eBOSS spectra~\citep{2020ApJS..249....3A}\footnote{\url{https://rcsed2.voxastro.org/data/galaxy/3038340}}.

\subsection{New spectroscopic observations}

We observed CARLA J0950+2743 with the 6.5m converted Multiple Mirror Telescope (MMT) using the MMT and Magellan InfraRed Spectrograph~\citep[MMIRS;][PI: I. Chilingarian, program SAO-18-23A]{2012PASP..124.1318M}.

To select galaxy targets, we first built a multi-wavelength catalog (IRAC1, IRAC2, and ACAM $i$-band) using SExtractor~\citep{1996A&AS..117..393B} in multiple detection mode, using IRAC1 as the detection image. Then, we selected galaxy candidates at $z>2$, applying a cut in the $(i-$IRAC1$)$ vs IRAC1 color--magnitude diagram following~\citet{2015MNRAS.452.2318C}.

Using the MMIRSMask web-tool\footnote{\url{https://scheduler.mmto.arizona.edu/MMIRSMask/index.php?}} we designed the two slit masks \textit{C0950p27} (hereafter \textit{mask1}) and \textit{C09p27\_2} (hereafter \textit{mask2}), each covering a rectangular area of the sky of 4.0$\times$6.9~arcmin with 6~arcsec long slitlets.

The slitlets were 0.8~arcsec, and 0.5~arcsec wide, for \textit{mask1} and \textit{mask2}, respectively, and matched the average seeing quality during observations in the $K_s$ band (0.7 and 0.55~arcsec, respectively).
We used the \emph{HK} grism with the \emph{HK3} cutoff filter that covers the wavelength range 1.25--2.35~$\mu$m (50\%\ transmission limits) at the spectral resolving power $R\sim1400$ and $R\sim 1700$ for the 0.8~arcsec and 0.5~arcsec slits, respectively. The selected setup covers AGN restframe H$\beta$ and [O{\sc iii}] emission lines in the \emph{H} band, and H$\alpha$ in the \emph{K} band. \textit{mask1} was observed on April 3, 6, and 8; and May 12, 2023, with a total integration time of 7~h. \textit{mask2} was observed on April 9 and 10, 2023, with a total integration time of 5~h. \textit{mask1} was also observed on May 7--8 2023 with 4~h of  total integration in the J/zJ setup, which covers the range $0.949 - 1.500$ $\mu$m (50\%\ transmission limits) at the spectral resolving power $R\sim 2200$ and includes the [O{\sc ii}] doublet at $z\sim2.3$.

Further details about spectroscopic observations and data reduction are provided in Appendix~\ref{ir_reduction}.

\begin{table*}
\centering
\caption{Spectroscopic measurements for confirmed cluster members of CARLA J0950+2743 field. Columns are: 1-2) Celestial coordinates, RA and Dec; 3) redshift estimate from H$\alpha$ and 4) its quality flag Q using the same notation as in \citet{2018ApJ...859...38N}; 5-7) measured aperture-corrected for PSF fluxes for [OII], [OIII] and H$\alpha$ emission lines and 8) the corresponding luminosity $L_{H\alpha}$ in H$\alpha$.}
\begin{tabular}{ c c c c c c c c}
\hline
RA (J2000) & Dec (J2000) & z & Q & $F_{[O\mathrm{II}]}$ & $F_{[O\mathrm{III}]}$ & $F_{H\alpha}$ & $L_{H\alpha}$ \\
deg. & deg. &  & & $10^{-17}$~erg~s$^{-1}$~cm$^{-2}$ & $10^{-17}$~erg~s$^{-1}$~cm$^{-2}$ &$10^{-17}$~erg~s$^{-1}$~cm$^{-2}$ & $10^{42}$~erg~s$^{-1}$ \\

\hline
\multicolumn{8}{c}{Galaxies with redshift close to the CARLA J0950+2743 AGN (z=2.36)}\\
\hline
147.64699 & 27.75788 & 2.3606 & A & 3.92 $\pm$ 0.25 & 3.72 $\pm$ 0.16 & 8.08 $\pm$ 0.45 & 3.31 $\pm$ 0.18 \\
147.63880 & 27.75036 & 2.3604 & A & 2.23 $\pm$ 0.22 & 6.90 $\pm$ 0.15 & 7.60 $\pm$ 0.44 & 3.12 $\pm$ 0.18 \\
147.64189 & 27.75431 & 2.3636 & A & 1.35 $\pm$ 0.18 & 3.82 $\pm$ 0.17 & 6.34 $\pm$ 0.33 & 2.60 $\pm$ 0.14 \\
147.64497 & 27.74475 & 2.3739 & A & 1.68 $\pm$ 0.26 & 2.46 $\pm$ 0.21 & 8.79 $\pm$ 0.37 & 3.61 $\pm$ 0.15 \\
147.65069 & 27.73948 & 2.3628 & A & 2.03 $\pm$ 0.21 & 1.76 $\pm$ 0.16 & 4.12 $\pm$ 0.32 & 1.69 $\pm$ 0.13 \\
147.64295 & 27.69930 & 2.3557 & B+&  &  & 4.86 $\pm$ 0.40 & 1.99 $\pm$ 0.17 \\
147.65038 & 27.69123 & 2.3620 & B+&  &  & 2.91 $\pm$ 0.35 & 1.19 $\pm$ 0.14 \\
147.62717 & 27.68929 & 2.3745 & A &  & 2.61 $\pm$ 0.38 & 5.32 $\pm$ 0.36 & 2.18 $\pm$ 0.15 \\
\hline
\multicolumn{8}{c}{Galaxies around z=2.24}\\
\hline
147.62799 & 27.73204 & 2.2420 & A & & 8.68 $\pm$ 0.54 & 7.06 $\pm$ 0.29 & 2.54 $\pm$ 0.10 \\
147.63791 & 27.72376 & 2.2450 & A & & 2.34 $\pm$ 0.29 & 5.84 $\pm$ 0.21 & 2.10 $\pm$ 0.08 \\
147.63577 & 27.70642 & 2.2567 & A & & 2.51 $\pm$ 0.20 & 3.53 $\pm$ 0.23 & 1.27 $\pm$ 0.08 \\
147.64754 & 27.71634 & 2.2450 & B+& &  & 3.41 $\pm$ 0.31 & 1.23 $\pm$ 0.11 \\
147.65736 & 27.74262 & 2.2286 & B+& &  & 4.92 $\pm$ 0.40 & 1.77 $\pm$ 0.14 \\
\hline

\end{tabular}
\label{tab:emlines}
\end{table*}

\subsection{Archival Chandra X-ray data}
\label{sec:xray_arch}
The CARLA J0950+2743 region was serendipitously observed by Chandra with the ACIS instrument on January 17, 2010 with an integration time of 8.2~ksec (dataset ID: 11376, PI: E. Gallo, target: PGC028305). This dataset unveils an  X-ray source close to the center of the galaxy overdensity (Fig.~\ref{fig_j0950}) in the ACIS-S0 detector, which is $\sim$15~arcmin away from the aimpoint. 

We measured the total flux of the X-ray counterpart as $6.77\pm1.51\times10^{-14}$~erg~s$^{-1}$~cm$^{-2}$ in the $0.5-5$~keV energy band in a circular aperture of radius 27~arcsec (220~kpc). The Analysis of its 1-D profile did not allow us to securely distinguish it from a point-source given a small number of photons. At the same time a Kolmogorov-Smirnov test of the observed distribution of photons shows that the data are consistent with a point source distribution only at the confidence level of $p=0.0037$, which confirms that X-ray the source has an extended component in addition to point source corresponding to the AGN. This is because this test is more sensitive to the signal 2-D distribution, e.g. to differences in ellipticity.
Further technical details about the analysis of the X-ray data are provided in Appendix~\ref{xray_analysis}. 

We discuss possible sources of the extended emission other than the ICM in Section~\ref{contamination}.
We conclude, that the observed X-ray source is likely a combination of the AGN and some extended component, and a more precise analysis of the source shape would require deep high-resolution observations.

\section{Results and Discussion}

\subsection{CARLA J0950+2743 spectroscopic confirmation and the serendipitous discovery of a new cluster at $z=2.24$}

We extract emission line fluxes using the optimal extraction method, described in \citet{1986PASP...98..609H}. For each individual emission line we modelled its 2-D profile with a single 2-D Gaussian, and  divided it by the error frame. We then extracted fluxes and uncertainties using a 2-D Gaussian weighting. Our measurements are shown in Fig.~\ref{fig_j0950_spec} and Table~\ref{tab:emlines}.

Following \citet{2018ApJ...859...38N}, we spectroscopically confirmed CARLA J0950+2743 using the \citet{2008ApJ...684..905E} criteria for $z > 1$ clusters of having at least 5 galaxies within $\pm2000 (1 + z)$~km~s$^{-1}$ from the AGN redshift range within a physical radius of 2~Mpc. In fact, we identified eight galaxies that satisfied these criteria with H$_\alpha$ detections ($SNR > 8$), of which six also show other prominent emission lines. The shorter J/zJ setup integration time led to lower SNR in the [O{\sc ii}] 3727$\AA$ detections, which might also be affected by stronger dust extinction. 
We measure a cluster redshift of $z=2.363\pm0.005$, as the average of the redshifts of all spectroscopically confirmed members.

We also discovered a foreground structure at $z=2.243\pm0.008$ (155~Mpc co-moving distance from CARLA J0950+2743), consistent with the \citet{2008ApJ...684..905E} criteria. This cluster, which we name CARLA-Ser J0950+2743 following \citet{2018ApJ...859...38N}, presents five galaxies with H$\alpha$ emission, and three that also show [OIII] emission (Fig.~\ref{fig_j0950_spec} and Table~\ref{tab:emlines}). Given that only three galaxies show multiple emission lines, this confirmation and cluster classification should be validated with further observations.

\subsection{Cluster mass constraints from X-ray data assuming a hot ICM}
\label{sec:clmass}

\begin{figure}
\centering

\includegraphics[width=\hsize]{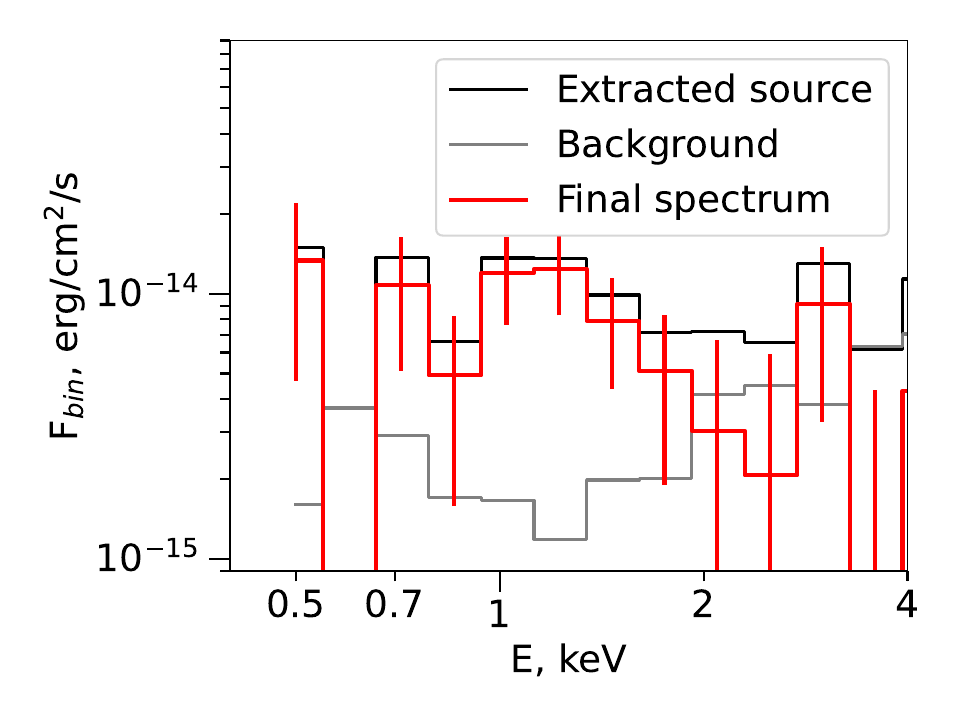}
\caption{The X-ray spectrum extracted in the aperture with a radius of 27 arcsec, centered on the AGN (black), and the spectrum of the background (grey). The red line shows the spectrum of the source with subtracted background.}
\label{fig_xray_spec}
\end{figure}

The estimated flux in the 0.5--5~keV (see Fig.~\ref{fig_xray_spec}) band corresponds to the observed luminosity of $L_{0.5-5 keV} = 2.9 \pm 0.6 \times10^{45}$~erg~s$^{-1}$. In this study we denote $L_{0.5-5 keV}$ as an X-ray luminosity in 0.5-5 keV observed band, which corresponds to 1.7-17 keV restframe at z=2.36.

Given the small number of detected photons, and hence, large errorbars, the shape of the X-ray spectrum is consistent with a wide range of possible components, including: a power-law component with $\Gamma=2$, typical for AGNs and all possible variations of ICM bremsstrahlung emission. We expect that the X-ray spectrum is likely to be a combination of these components. However, the limited depth of the dataset prevents us from a more precise decomposition. Hereafter, we choose a bremsstrahlung model for the parametrization of the X-ray spectral shape, but we verified that the choice of other models, or their combination, does not have a significant effect on the results. The use of pure bremsstrahlung model delivers the most  conservative estimate of k-correction range given that for the power-law with $\Gamma=2$ k-correction is 1.

To estimate the k-correction, we modeled the observed X-ray spectrum with {\sc XSpec}~\citep{2001ASPC..238..415D} using the model {\sc apec} that corresponds to ``free-free'' transitions, which is a dominating regime of emission in sparse hot plasma in galaxy clusters~\citep{2010A&ARv..18..127B}. This modelling resulted in an electron temperature estimate of $T_X = 5.77_{-1.09}^{+10.2}$~keV. This temperature corresponds to a $k$-correction of $k_X=0.612$ that is needed to convert the observed flux into the restframe. Using the {\sc Sherpa} tool~\citep{2007ASPC..376..543D}, we estimate $L_{0.2-2 keV} = 1.7 \pm 0.4 \times10^{45}$~erg~s$^{-1}$ and $L_{X} = 4.5 \pm 1.0 \times10^{45}$~erg~s$^{-1}$\footnote{Hereafter, with $L_X$ we denote ``bolometric'' luminosity following~\citet{2014ApJ...794...48C}}. 

If we assume a redshift of $z=2.363$, according to the $L_X$--$M_{2500}$ relation~\citep{2014ApJ...794...48C},  the luminosity estimate corresponds to  a total dark matter mass $M_{2500} = 0.85 \pm 0.07_{(stat.)} \pm 0.26_{(sys.)} \times 10^{14} M_{\odot}$, considering the $1\sigma$ scatter in the relation, which is 0.1~dex (added to the systematic uncertainty). We also estimate the mass iteratively, using two scaling relations -- the M-T and the T-L relation; this allows to avoid uncertainties related to k-corrections, given that we don't need to estimate the k-correction from observations, but only from the temperature using M-T scaling relation. This approach yielded $M_{2500} = 0.80 \pm 0.25 \times 10^{14} M_{\odot}$, which is very close to the mass estimate that uses k-corrections. 

Considering the transformations calibrated on the Magneticum cosmological hydrodynamic simulation~\footnote{\url{https://c2papcosmosim.uc.lrz.de/static/hydro_mc/webapp/index.html}}~\citep{2021MNRAS.500.5056R}, we converted the measurement of $M_{2500}$ to $M_{200}=3.30 ^{+0.23}_{-0.26 (\mathrm{stat})}$ $^{+1.28}_{-0.96 (\mathrm{sys})} \times10^{14} M_{\odot}$. If using a redshift $z=2.243$, our mass estimate does not change significantly provided that the difference of the luminosity distances results in a luminosity ratio of 1.136. In fact, given the power-law index of the $L_{X}-M_{2500}$ relation of 0.305, the mass estimate would be 8\%\ lower. 

We use $T_X = 5.77_{-1.09}^{+10.2}$~keV to estimate the $k$-correction, therefore its uncertainty contributes to the systematic uncertainty of the mass estimate. In fact, for the 3-$\sigma$ lower limit of the temperature ($T_X=2.50$~keV), the $k$-correction is $k_X=1.43$ yielding $M_{200}= 3.76\times10^{14} M_{\odot}$. At the same time, the $3\sigma$ upper limit for $T_X$ of 35~keV is not physically possible, because the highest values observed in galaxy clusters are of the order of $T_X\sim10$~keV yielding $k_X=0.414$ that corresponds to $M_{200}=3.25\times10^{14} M_{\odot}$. 

Considering both statistical and systematic errors, which are independent, we conclude that,  if all the X-ray emission would be due to an extended ICM emission, it corresponds to a total dark matter mass of $M_{200} \approx 3.0-3.3^{+0.23}_{-0.26 (\mathrm{stat})}$ $^{+1.28}_{-0.96 (\mathrm{sys})}$ $\times10^{14} M_{\odot}$ for the two clusters. 

Since our observations do not permit us to confirm a thermal emission, a discussion about alternative explanations for the observed X-ray extended contribution is given in the Appendix~\ref{contamination}.

We also have to take into account that at least part of the X-ray emission is due to the AGN. The total X-ray luminosity and the estimated AGN luminosity from multi-wavelength data are consistent within $\sim$3~$\sigma$ (see Appendix~\ref{agn_contrib}). This means that we cannot formally rule out that the X-ray flux is completely due to the AGN. 

However, if that were true, it would contradict our results from the KS-test that show that the observed X-ray emission is not consistent with a point source and, moreover, the X-ray emission is not centered on the AGN but rather on the galaxy overdensity. 

Our best estimate of the AGN contribution from multi-wavelength scaling relations of the X-ray surface brightness distribution is 32$\pm$22\% (Appendix~\ref{est_agn_joint}), and yield our final total dark matter mass estimate of $M_{200} \approx 2.7-3.0^{+0.20}_{-0.23 (\mathrm{stat})} $ $^{+1.13}_{-0.85 (\mathrm{sys})}$ $\times10^{14} M_{\odot}$. However, the scatter in the fraction of a possible AGN contribution does not have a very strong affect on the mass estimate: in a very pessimistic case of a +2$\sigma$ outlier resulting in an AGN contribution of 76\%, the mass estimate will change only to $M_{200} \approx 2.0-2.2^{+0.14}_{-0.16 (\mathrm{stat})} $ $^{+0.83}_{-0.62 (\mathrm{sys})}$ $\times10^{14} M_{\odot}$.

\subsection{Perspectives of the cosmological implications}

CARLA J0950+2743/CARLA-Ser~J0950+2743 and their X-ray counterpart can be used for future studies of structure formation in the cosmological context. However, the interpretation of the X-ray observations of these clusters will be also affected by the superposition of these systems. 

If the diffuse emission originates from the ICM in only one of these two clusters, its total luminosity can be used to put a lower limit on the cluster mass, given that the luminosity of a more massive cluster is always higher (or equal) than the sum of the luminosities of two clusters.

\begin{figure}
\centering

\includegraphics[width=\hsize]{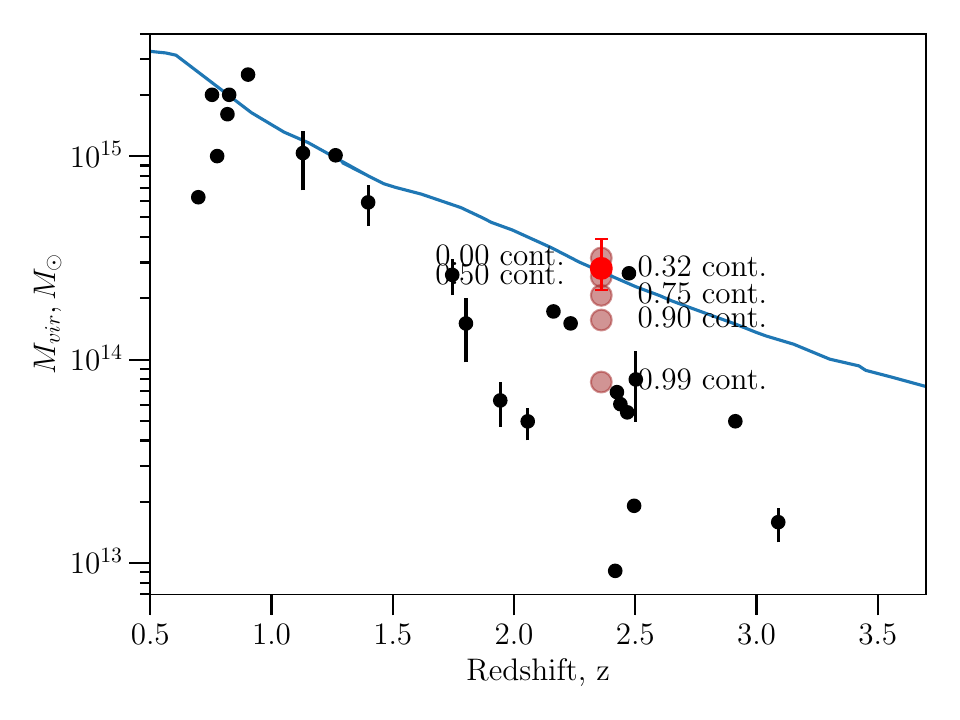}

\caption{Comparison of CARLA J0950+2743 mass estimate under assumptions of different contamination levels from 0 to 99\% (brown points), including the most realistic value of 32\% (red point) with measurements for other structures (black points), compiled in \citet{2023ApJ...950..191R}, that includes clusters from \citet{2003ApJ...586L.111S, 2004A&A...424L..17V, 2005ApJ...620L...1O, 2012ApJ...750..137T, 2014ApJ...792...15T, 2019ApJ...879...28H, 2019ApJ...877...51C, 2020ApJ...888...89T, 2021MNRAS.502.4558C} and virial mass of the most massive halo in the snapshots at different redshifts (blue line) for ``Box 0'' simulation set.}
\label{fig_Mvir_z}
\end{figure}

The analysis of the Magneticum Pathfinder suite of cosmological simulations shows that the virial masses of clusters at $z=2.36$ in case of the largest ``Box~0'' (2688~Mpc~h)$^{-3}$ box size~\citep{2023ApJ...950..191R} do not exceed $M_{vir}\simeq3\times10^{14} M_{\odot}$. This value is close to the total dark matter mass estimate for our two clusters derived from the X-ray analysis, when we convert $M_{200}$  to $M_{vir}$ following \citealp{2021MNRAS.500.5056R}. We show our two cluster total virial mass in color in Fig.~\ref{fig_Mvir_z}, considering different percentage levels of contamination from the AGN. 

Even under the assumption of an AGN X-ray contamination level of 90\%\ (e.g., $\approx$ 2.5-$\sigma$ higher that the average contamination derived from scaling relations), if the X-ray thermal emission were dominated by one of the two clusters, it would still be among the most massive cluster at its redshift.

The implications of CARLA J0950+2743/CARLA-Ser~J0950+2743 in the cosmological context require deeper high-resolution X-ray observations to precisely constrain contribution from AGN and a possible contamination from other point sources, like those found in the field of some other high-redshift galaxy clusters and proto-clusters such as Spiderweb at $z=2.16$ \citep{2022A&A...662A..54T}, and deeper and more complete spectroscopic observations.

\section{Summary}
Using deep ground-based NIR spectroscopy, we confirmed the CARLA J0950+2743 galaxy cluster at $z=2.363\pm0.005$ following the \citet{2008ApJ...684..905E} criteria, i.e., by confirming eight spectroscopic members within 2000$\times(1+z)$~km~s$^{-1}$ and a physical radius of 2~Mpc, including five galaxies with multiple emission lines.

We also surreptitiously discover another structure, CARLA-Ser~J0950+2743 at $z=2.243\pm0.008$, that we classify as a cluster following the \citet{2008ApJ...684..905E} criteria. However, this last classification is based on five spectroscopically confirmed members, of which only three with multiple emission lines. This confirmation would benefit from deeper observations to obtain more galaxies with multiple emission lines, while the \citet{2008ApJ...684..905E} criteria would need at least five sure spectroscopically confirmed members. 

Archival X-ray observations by the Chandra observatory reveal a counterpart that is consistent with the existence of a total dark matter halo of mass $M_{200} \approx 2.7-3.0^{+0.20}_{-0.23 (\mathrm{stat})} $ $^{+1.13}_{-0.85 (\mathrm{sys})}$ $\times10^{14} M_{\odot}$  for the two clusters, if the emission is associated with the cluster ICM. 

To improve each cluster mass estimate, deeper high-resolution X-ray observations are needed to better constrain the AGN contribution to the signal.

CARLA J0950+2743 and CARLA-Ser J0950+2743 provide an unique opportunity to study the formation of the large scale structure at the age of the Universe of 2--3 billion years after the Big Bang, as well as evolution of galaxy population in dense environments and the main factors driving it.

\begin{acknowledgements}
We thank Universit\'e Paris Cité, which founded KG's Ph.D. research. We also thank Franz Bauer and Alexei Vikhlinin for useful discussions. KG thanks Victoria Toptun for the fruitful discussions related to the X-ray analysis. IC's research is supported by the Telescope Data Center, Smithsonian Astrophysical Observatory. Observations reported here were obtained at the MMT Observatory, a joint facility of the Smithsonian Institution and the University of Arizona. We gratefully acknowledge support from the CNRS/IN2P3 Computing Center (Lyon - France) for providing computing and data-processing resources
needed for this work.
\end{acknowledgements}

\bibliographystyle{aa}
\bibliography{references.bib}

\begin{appendix}

\section{Spectroscopic observations and data reduction}
\label{ir_reduction}
For the MMIRS observations, we used a 4-position dithering pattern (ABA'B') at $+1.4, -1.0, +1.0, -1.4$~arcsec. The individual exposure times were set to 300~sec with the 4.426~sec up-the-ramp non-destructive readout sequence using the 0.95~e-/ADU inverse gain. The readout noise per readout $\sim$15~e- was reduced to the effective value of $\sim$3~e- after 69 readouts. 

We reduced data with the MMIRS pipeline~\citep{2015PASP..127..406C}, which included the following steps: (i) reference pixel correction and up-the-ramp fitting of raw readouts; (ii) dark subtraction; (iii) flat fielding; (iv) extraction of 2D slitlets; (v) wavelength solution using OH lines; (vi) sky background subtraction using a modified \citet{2003PASP..115..688K} technique with a global sky model; (vii) correction for the telluric absorption and relative flux calibration using observations of a A0V telluric standard star. We ran the pipeline on individual A$-$B (or A'$-$B') dithered pairs.

Then, we co-added the dithered pairs from observations collected during different nights applying the weights inversely proportional to the squared seeing FWHM. 

At the end, we performed the absolute flux calibration by using secondary calibration stars included in the masks by re-normalizing their fluxes to the $H$ and $K_s$-band measurements from the UKIRT Hemisphere Survey \citep[UHS][]{2018MNRAS.473.5113D}.

Our datasets reach a 3$\sigma$ sensitivity at  $1.1\times10^{-17}$~erg~s$^{-1}$~cm$^{-2}$ and $1.4\times10^{-17}$~erg~s$^{-1}$~cm$^{-2}$, for \textit{mask1} and \textit{mask2}, respectively, for a typical $H_{\alpha}$ emission line, with the restframe full width at half maximum of 9.7~$\AA$. This corresponds to a  $1.2\times10^{-18}$~erg~s$^{-1}$~cm$^{-2}$~\AA$^{-1}$ and $1.5\times10^{-18}$~erg~s$^{-1}$~cm$^{-2}$~\AA$^{-1}$, for \textit{mask1} and \textit{mask2}, respectively, in the continuum averaged between H$\alpha$ and [S{\sc ii}].

For \textit{mask1}, the achieved depth in J/zJ reached $1.6\times10^{-17}$~erg~s$^{-1}$~cm$^{-2}$ for the [O{\sc ii}] emission line, and $1.8\times10^{-18}$~erg~s$^{-1}$~cm$^{-2}$~\AA$^{-1}$ for the continuum in the region of this emission line.

\section{Spatial extent of an X-ray counterpart}
\label{xray_analysis}

We measured the flux of the X-ray cluster counterpart in the following way: (i) We selected a subsample of the registered events within a radius of 27~arcsec from the peak of the counts in the binned dataset, which identify our target; (ii) To estimate the background, we select a subsample of events in two circular areas with radii 86 and 62~arcsec, in the same detector, but far enough from the extraction region of the main source; (iii) Then we calculated the energy spectra of these three photon subsamples; (iv) We subtracted the background from the target spectrum, after renormalising by the area covered by the regions; (v) We corrected the spectra taking into account the photon energy and effective area. We obtained a total target flux of $6.77\pm1.51\times10^{-14}$~erg~s$^{-1}$~cm$^{-2}$ in the $0.5-5$~keV energy band in the circular aperture with a radius of 27~arcsec (220~kpc).

Using the {\sc ChaRT} web-tool\footnote{\url{https://cxc.cfa.harvard.edu/ciao/PSFs/chart2/runchart.html}} we ran a raytrace simulation for the position of the AGN on the archival Chandra dataset for 1000 rays for a source with a spectral shape obtained from the modelling with {\sc XSpec} as discussed in Section~\ref{sec:clmass}. The obtained ray map was then reprojected on the detector plane using {\sc MARX}~\citep{2012SPIE.8443E..1AD} to obtain a model of the Chandra Point Spread Function (PSF).

\begin{figure}
\centering
\includegraphics[width=\hsize]{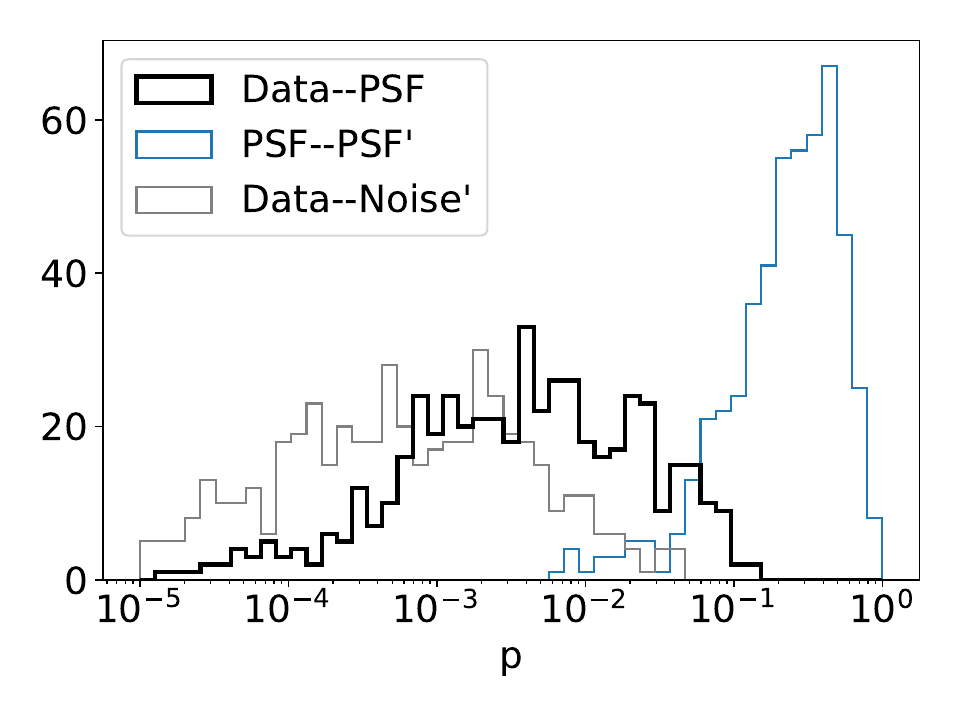}
\caption{The distribution of the significance levels (p) for different realisations of the mock event lists obtained using the KS test for the observed distribution  vs a point source distribution (black). As an additional test, we present the distribution of p for the cross-tested generated samples of event lists that correspond to point-source (blue) and for the tested background-only photon distribution (grey).}
\label{fig_xray_kstest}
\end{figure}

\begin{figure}
\centering
\includegraphics[width=\hsize]{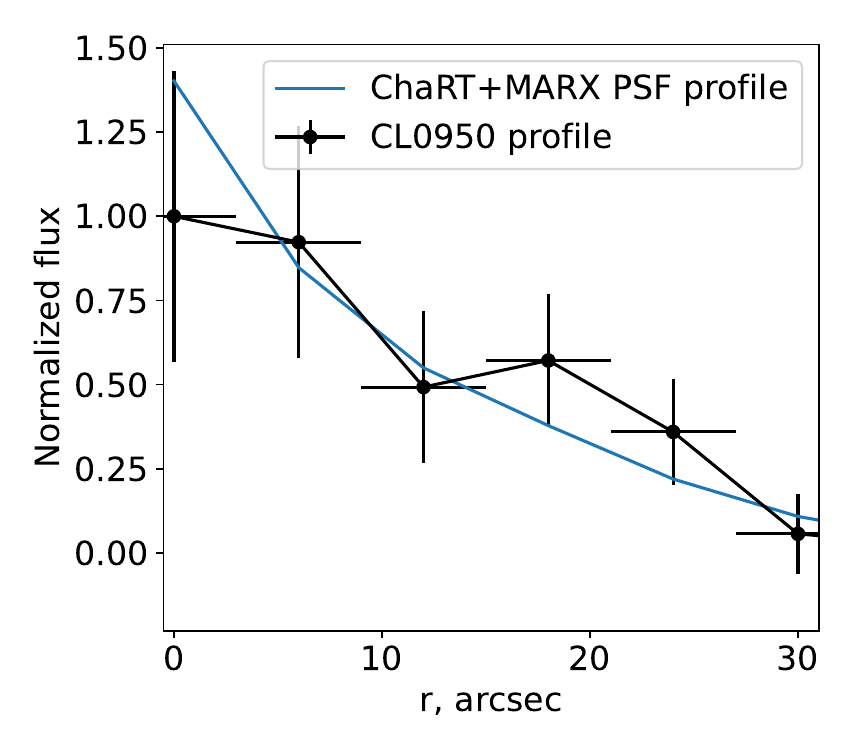}
\caption{Radial profile of detected photons extracted from the Chandra dataset in circular annuli centered on the CARLA J0950+2743 AGN (black), compared to the PSF profile modelled using ChaRT+MARX (blue) with the normalisation factor derived from the $\chi^2$ minimization. }
\label{fig_xray_psf_prof}
\end{figure}

\begin{figure}
\centering
\includegraphics[width=\hsize]{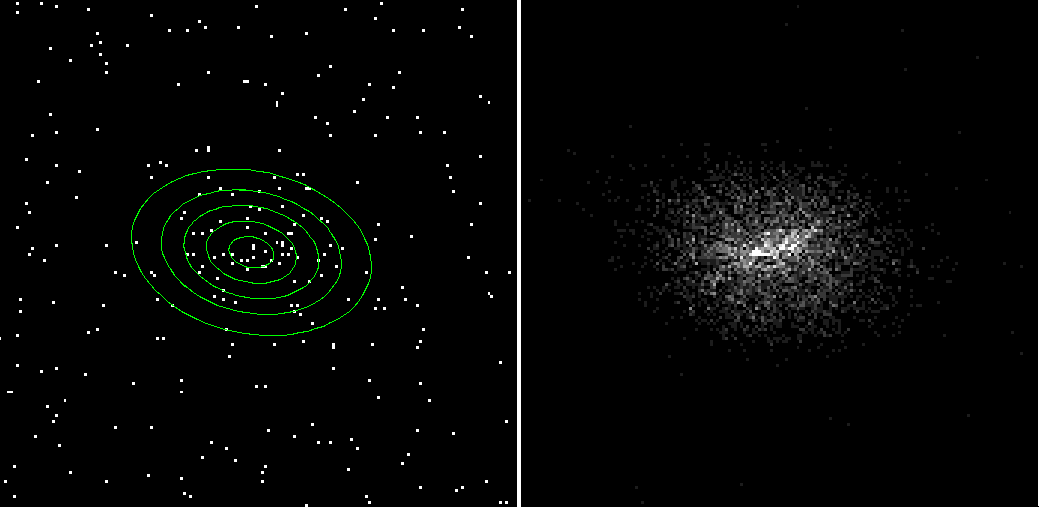}
\includegraphics[width=\hsize]{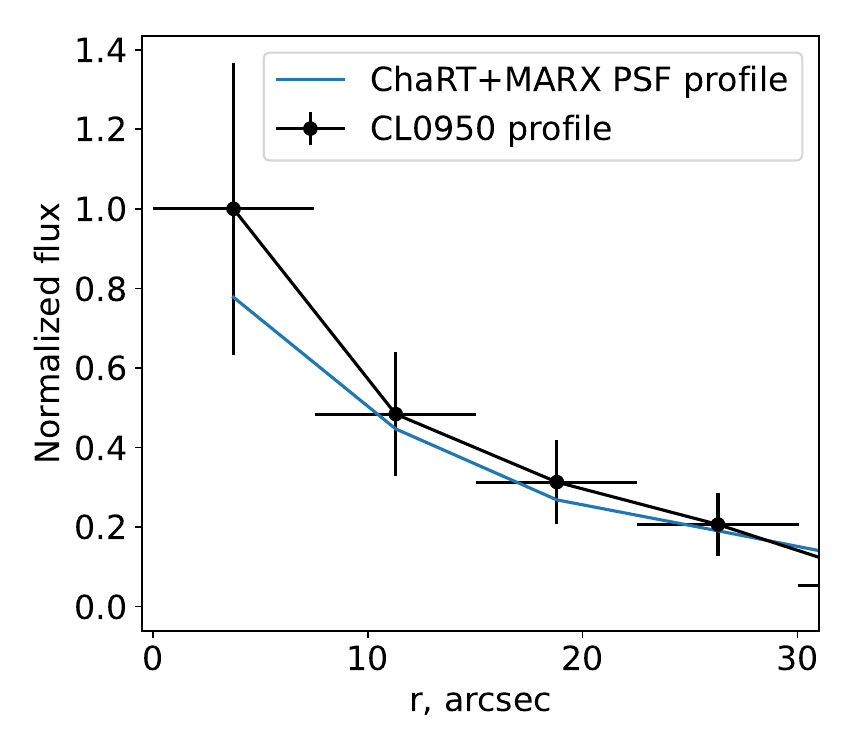}
\caption{Upper panel: Chandra image of CARLA J0950+2743 (left) and simulated PSF (right) with overlapped the elliptical annuli used for the 1-D signal profile (in green). 
All regions are centered on the AGN.
Lower panel: Radial profile of the detected photons per unit area in the Chandra dataset centering on the CARLA  J0950+2743 AGN, compared to the simulate Chandra PSF radial profile (in black). Both profiles are normalized to the total number of net counts found in the extraction regions (see upper panel).}
\label{fig_xray_psf_prof2}
\end{figure}

To asses whether the distribution of detected photons corresponds to what is expected for a point source, we performed a Kolmogorov--Smirnov test (KS-test)~\citep{Kolmogorov1933, Smirnov1948}. For this test, we estimated a 2-D distribution of detected photons that contains two components: background and a point source. To estimate the contribution of each component, we spatially binned the observed photon events within a square window with the side of 600 pixels and the bin size of 32 pixels. Then, we modelled the derived 2-D distribution by a linear combination of a constant component (background) and a PSF model, inferred coefficients were used to assign background and point source the same flux amplitude as in the Chandra dataset.

We use a classical two-sample KS test~\citep{1983MNRAS.202..615P, 1987MNRAS.225..155F, 2002nrca.book.....P}, which provides a better ability of separating two distributions. 

Using the 2DKS Python package\footnote{\url{https://github.com/Gabinou/2DKS}}, we ran a KS test for the observed distribution of detected photons and 500 realisations of the model distribution. The median value of the KS-statistic D (the maximum deviation between the observed cumulative distribution function of the sample and the cumulative distribution function of the model distribution) is $D=0.0642$, with a median value of the significance level to follow the same distribution of $p=0.0037$ (see Fig.~\ref{fig_xray_kstest}). 

As an additional cross-validation, we ran the same KS test for independently generated mock event lists that follow the model distribution. This test yielded a median $D=0.0362$ and $p=0.27$. At the same time to confirm the source detection, we also did a similar test between observed photon distribution and a pure background distribution what yielded a median $D=0.0735$ and $p=0.00054$, which securely confirms the presence of a source. From these results we conclude that: (i) Using a two-sample KS test for the Chandra dataset we can separate with high confidence a case of a pure point-source from a combination of a point source and extended one; (ii) The observed observed photon distribution cannot be described by a single point source and constant background at a significance level of 0.997.

In Figures~\ref{fig_xray_psf_prof} and~\ref{fig_xray_psf_prof2}, we show the radial profiles of the cluster counterpart in X-ray obtained with 55 photons and a generated PSF model.
In Figure~\ref{fig_xray_psf_prof} and~\ref{fig_xray_psf_prof2} we provide the 1-D profile for the source and for PSF model extracted in circular and elliptical (b/a=0.75) annuli respectively, both centered on the position of the AGN. The profile derived with elliptical annuli is normalized to the total number of photons, while normalization of the coefficients for the profile in circular annuli is obtained by using a $\chi^2$ minimization.

Both 1-D profile analyses are consistent with emission from a point source, but the extended emission detected from our spatial analysis is also consistent with these profiles, given the large uncertainties on the data.

The spatial distribution of our source is also substantially rounder ($b/a=0.80\pm$0.16) than the PSF ($b/a=0.5$) at that position in the Chandra FoV (Fig.~\ref{fig_xray_decomp}), and this could explain why our spatial analysis and KS test are more conclusive to show evidence for an extended source component.

The error bars clearly demonstrate that our modelling of the observed photon distribution is limited by the statistical effects, i.e. number of detected photons, rather than the PSF systematics.

\section{Estimates of the AGN contribution to X-ray luminosity}
\label{agn_contrib}

\subsection{Estimates of the AGN X-ray luminosity}
\label{est_agn_joint}
The only available archival Chandra X-ray dataset does not provide a precise measurement of the AGN contribution, therefore we use several multi-wavelength scaling relations to estimate it. Each of these relations has a relatively high intrinsic spread, however, because they probe different physical regions and/or mechanisms of an AGN, together they provide a very good constraint of the flux.

By combining the estimates obtained using relations between AGN luminosities in the X-ray and UV or IR, we estimate a possible AGN contribution as $L_{AGN} / L_{tot} = 0.32_{-0.12}^{+0.22}$, taking the average of all contributions estimated in the following subsection, Fundamental Plane of the black hole activity for which we cannot conclude with reasonable uncertainties. For the uncertainties, we took into account the scatter of the scaling relations, the measurements of the X-ray source flux, and the uncertainty on the variables used in the scaling relations. The details of our estimations are given in the subsections below.

\subsubsection{Fundamental Plane of the black hole activity.}
The fundamental plane of black hole activity \citep{2003MNRAS.345.1057M} relates the black hole mass, the radio luminosity at 5~GHz, and the X-ray luminosity in the 2--10~keV range. It suggests that the processes governing black hole accretion and jet emission are scalable and can be described by universal laws for black holes in the mass ranges from stellar to supermassive. A recent re-calibration of the fundamental plane of black hole activity based on observational data for intermediate-resolution quasars \citep{2022MNRAS.513.4673B} has a scatter of $\sim$0.62~dex for radio-loud AGN, which is smaller than that of the original relation \citep[][$\sim$0.8-0.9~dex]{2003MNRAS.345.1057M}. But it still leads to $1\sigma$ uncertainties of a factor of $\sim$3 in X-ray luminosity. The black hole mass estimate for the AGN from the broad lines (C{\sc iv}]) is $M_{BH} = 2.0-2.6 \times 10^9 M_{\odot}$~\citep{2012ApJ...753..106M, 2017ApJS..228....9K} and the radio continuum luminosity $\nu F_{\nu}=2.9\times10^{44}~erg~s^{-1}$ estimated from \citet{2010A&A...511A..53V} yields the AGN X-ray luminosity of $L_{0.5-5 keV} = 3.79_{-2.88}^{+12.0}\times 10^{46}erg/s$, which substantially exceeds the X-ray luminosity that we measured for our source. However, given the large scatter, we cannot consider this estimate for assessing the AGN contribution, but can only conclude that it can be between 20\% and 100\% within the $3\sigma$ range. 

\subsubsection{$L_{0.2-2 keV} - L_{6\mu m}$ relation.}
The relation between the AGN luminosity in the X-ray and in  the infrared (IR) reflects the tight dependence of the emission of the hot corona and the emission from the accretion disk irradiated by the dust~\citep{2004A&A...418..465L, 2009A&A...502..457G, 2009A&A...498...67L}. 
 We use the correlation between the mid-IR flux at 6~$\mu$m \citep{2015ApJ...807..129S} and X-ray that relates the X-ray emission to the emission of the warm dust in the AGN torus, and is less affected by the intrinsic dust attenuation unlike the correlations based on the UV luminosity. Using the WISE W4 magnitude measurement for the AGN that perfectly matches the restframe 6 ~$\mu$m, we estimated $L_{0.5-5 keV} = 0.75\times 10^{45}$~erg~s$^{-1}$, or $\sim$25\% of the total observed X-ray luminosity. 

The scatter, estimated for the "filtered high-luminosity quasar" sample from~\citet{2007ApJ...665.1004J}, used to build the regression fit in \citet{2015ApJ...807..129S} is 0.26~dex.  Adding the uncertainty of the W4 magnitude measurement, this leads to the final estimate of $L_{0.5-5 keV} = 0.75_{-0.39}^{+0.71}\times 10^{45}$~erg~s$^{-1}$.  This corresponds to an AGN contribution of $0.26_{-0.13}^{+0.22}$.

\subsubsection{$L_{2keV} - L_{NUV}$ correlation.}
We use the empirical correlation between the AGN luminosity in soft X-ray and the near-UV from \citet{2010A&A...512A..34L} with an intrinsic scatter of 0.35~dex. It relates the accretion disk luminosity in the continuum at 2500\AA\ to the X-ray generated by the corona.

From available SDSS spectra, at $\lambda=8400$~\AA, we can directly measure the UV restframe continuum flux as $F_{2500A}$ = $3\times10^{-17}$~erg~s$^{-1}$~cm$^2$~\AA$^{-1}$ that corresponds to the UV restframe luminosity density of $L_{\nu, 2500A} = 1.10\times10^{31}$~erg~s$^{-1}$~cm$^2$~Hz$^{-1}$. 

From this, we derive the X-ray luminosity spectral density $L_{2keV} = 5.1\times10^{26}$~erg~s$^{-1}$~cm$^2$~Hz$^{-1}$~\citep{2023MNRAS.522.1247K}.  The uncertainty on $F_{2500A}$ is less than 5\%, so it does not affect the uncertainty of $L_{2keV}$ estimate. 

Using the {\sc calc\_kcorr} procedure in the Chandra {\sc Sherpa} toolkit, we converted the $L_{2keV}$ to the observed X-ray luminosity in the $0.5-5$~keV bandpass and obtain $L_{0.5-5 keV} = 1.1\times10^{45}$~erg~s$^{-1}$ assuming $\Gamma = 1.4$ for RLAGN ~\citep{1999ApJ...526...60S, 2011ApJ...738...53S}. Adding the scatter of the $L_X-L_{2500\AA}$ correlation of 0.35~dex, we obtain $L_{0.5-5 keV} = 1.1_{-0.6}^{+1.2}\times10^{45}$~erg~s$^{-1}$.
This corresponds to an AGN contribution of $0.38_{-0.22}^{+0.42}$.

\subsection{AGN contribution from the decomposition of the X-ray photon distribution}

\begin{figure*}
\centering
\includegraphics[width=0.8\hsize]{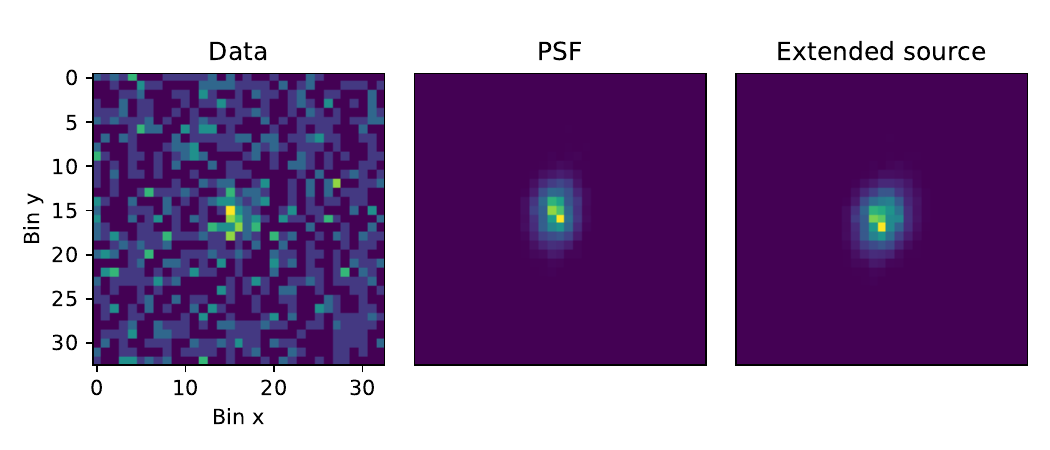}
\caption{Decomposition of a binned Chandra X-ray event list for CARLA J0950+2743. \textit{Left:} A binned event list with higher energy cut of 7keV; \textit{Middle}: A binned event list for the point source; \textit{Right}: A binned 2-D Gaussian model convolved with the Chandra PSF for the extended component. }
\label{fig_xray_decomp}
\end{figure*}

To directly constrain a contribution from the AGN point source in X-ray, we performed a decomposition of the observed background subtracted distribution of detected X-ray photons by representing it with a linear combination of a point-source component used in Section~\ref{sec:xray_arch}, and an extended Gaussian distribution, both convolved with a PSF (Fig.~\ref{fig_xray_decomp}). For a bin size of $18\times18$ pixels (9~arcsec)\footnote{We can consider this bin size optimal because it yields SNR$>$3 in most bins in the area of the detection and it does not yet affect much the spatial information.}, the recovered contribution of the point source is 35$\pm$26\%.  

Given that the shape of the point source component is fixed while the shape of the diffuse emission is flexible, the estimated AGN fraction can be treated as a upper limit of AGN fraction in the observed X-ray luminosity.

\subsection{Extended X-ray emission from inverse Compton scattering on a radio jet}
\label{contamination}

The observed X-ray counterpart can be explained by sources other than the ICM. For example, deep high-resolution Chandra X-ray imaging of the Spiderweb proto-cluster revealed a population of point sources (AGN)~\citep{2022A&A...662A..54T} that would be interpreted as extended ICM emission in case of lower angular resolution, similar to the Chandra dataset for our cluster. 

The inverse Compton (IC) scattering of the CMB photons on the electrons in the AGN jet can produce substantial X-ray emission~\citep{1979MNRAS.188...25H}. To estimate the possibility of high contribution of possible IC from the AGN to the observed X-ray luminosity, we followed an approach similar to that used for the cluster J1001+02 at $z=2.51$~\citep{2016ApJ...828...56W}, assuming that all the observed radio and X-ray emission originates from the AGN. 

Our calculations show a magnetic field estimate of $\sim$0.5$\mu$G that is almost an order of a magnitude smaller than typical magnetic fields in the systems where the IC is observed~\citep{2005A&A...433...87O}. We therefore conclude that the observed extended X-ray emission is unlikely to be produced by IC scattering.

The observed X-ray flux can also emerge from the IC structures in the radio-lobes~\citep{2006MNRAS.371...29E}, but observed X-ray luminosities of these objects are of the order of a few times of $10^{44}$ erg/s, substantially lower then expected luminosity of the diffuse emission. However, the very existence of lobe may already serve as a indirect confirmation of the ICM.

\end{appendix}

\end{document}